\documentstyle[12pt,epsfig]{article}

\newcommand{\be}{\begin{equation}}
\newcommand{\ee}{\end{equation}}
\newcommand{\ba}{\begin{eqnarray}}
\newcommand{\ea}{\end{eqnarray}}
\newcommand{\bb}{\begin{thebibliography}}
\newcommand{\eb}{\end{thebibliography}}
\newcommand{\ci}[1]{\cite{#1}}
\newcommand{\bi}[1]{\bibitem{#1}}
\newcommand{\lab}[1]{\label{#1}}

\begin{document}
\sloppy
\thispagestyle{empty}

\mbox{}
\vspace*{\fill}
\begin{center}
{\LARGE\bf  Study of Single-Spin Asymmetry in } \\

\vspace{2mm}
{\LARGE\bf Diffractive High- $p_t$- Jet
Production.}\\

\vspace{2em}
\large
S.V.Goloskokov
\footnote{Email: goloskkv@thsun1.jinr.dubna.su}
\\
\vspace{2em}

{
\it
Bogoliubov Laboratory of Theoretical  Physics,
  Joint Institute for Nuclear Research.}
 \\

{
\it
Dubna 141980, Moscow region, Russia.}\\
\end{center}
\vspace*{\fill}
\begin{abstract}
\noindent
It is shown that the transverse
single spin asymmetry in polarized diffractive
$Q \bar Q$ production depends strongly on the spin structure
of the pomeron coupling.
It is concluded that the spin properties of quark-pomeron and
proton-pomeron vertices can be studied in future HERA-N experiments.
\end{abstract}
\vspace*{\fill}
\newpage
%

\section{Introduction}
\label{sect1}
It is well known that large spin
asymmetries are observed in different high-energy reactions.
Extensive information about spin-dependent
distributions inside a hadron can be obtained from double spin asymmetries
\ci{soff}.
However, it is necessary to have two polarized particle beams (or a polarized
beam and a target) to study such asymmetries. For hadron high-energy reactions
it will be possible in RHIC.

A very important information on spin properties of QCD can be obtained
>from transverse single spin asymmetry. Single spin asymmetry differs from
double spin asymmetries and depends strongly on
the hadron--wave--function properties and is determined by the relation
\be
A=\frac{\sigma(^{\uparrow})-\sigma(^{\downarrow})}
{\sigma(^{\uparrow})+\sigma(^{\downarrow})} \propto
 \frac{\Im (f_{+}^{*} f_{-})}{|f_{+}|^2 +|f_{-}|^2},
\ee
where $f_{+}$ and $f_{-}$ are spin-non-flip and spin-flip amplitudes
respectively. So, single spin asymmetry appears if both $f_{+}$ and $f_{-}$
are non-zero and there is a phase shift between these amplitudes.
It can be shown that the spin-flip amplitude is of the order of magnitude
$$
f_{-} \propto \frac{m}{\sqrt{P_t^2}} f_{+}.
$$
Moreover, an additional loop in the amplitude is important to have a
phase shift between amplitudes.
As a result, we have
\be
A \propto \frac{m \alpha_s}{\sqrt{P_t^2}}.  \lab{aasy}
\ee
It has been shown in  \ci{ter} that the mass $m$ in (\ref{aasy}) is of the
order of the hadron mass. So, we can expect large transverse asymmetry for
$P_t^2 \simeq {\rm Few}~ GeV^2$. For such momenta transfer the diffractive
processes should be important.

Diffractive production of high $p_t$ jets has been observed experimentally
at CERN and HERA in hadron-hadron collisions  and deep inelastic
lepton-proton scattering \ci{gap}. Such processes where a proton remains
intact are determined at high energies by the pomeron
exchange. These experiments have been initiating various
investigations of the
diffractive reactions and pomeron properties.

The pomeron is a colour--singlet vacuum $t$-channel exchange that can be
regarded as a two-gluon state.
The pomeron contribution to the hadron high-energy amplitude can be written
as a product of two pomeron-hadron vertices $V_{\mu}^{hhI\hspace{-1.1mm}P}$
multiplied by some function $I\hspace{-1.6mm}P$
determined by the pomeron.
As a result,  the quark-proton high-energy amplitude
looks as follows
\be
T(s,t)=i I\hspace{-1.6mm}P(s,t) V_{qqI\hspace{-1.1mm}P}^{\mu} \otimes
V^{ppI\hspace{-1.1mm}P}_{\mu}.    \lab{tpom}
\ee

In the nonperturbative two-gluon exchange model  \ci{la-na} and the
BFKL model \ci{bfkl} the pomeron couplings have a simple matrix
structure:
\be
V^{\mu}_{hh I\hspace{-1.1mm}P} =\beta_{hh I\hspace{-1.1mm}P}\; \gamma^{\mu},
\lab{pmu}
\ee
 which leads to  very small spin-flip effects. We shall call this form the
 standard coupling.

However, the experimental study of transverse spin asymmetries in diffractive
reactions shows that
at a high energy and momentum transfer $|t| >1 GeV^2$ they are not
small \ci{nur} and can possess a weak energy dependence.
This means that the pomeron can be complicated in the spin structure
\ci{zpc,akch}.

The structure of the pomeron-proton coupling  is significant in the
explanation
of transverse asymmetries. Information about this coupling is rather
limited. Model approaches show that the pomeron--proton vertex is of the form
\be
V_{ppI\hspace{-1.1mm}P}^{\mu}(p,r)=m p_{\mu} A(r)+ \gamma_{\mu} B(r).
\lab{prver}
\ee
The amplitudes $A$ and $B$ in (\ref{prver}) are determined by the proton wave
function and can be calculated within models approaches
(see  e.g. \ci{zpc,kop}).
For spin-averaged and longitudinal polarization of the proton,
the $B$ term is predominant. As a result, the longitudinal double spin
asymmetry does not depend on the pomeron-proton vertex structure.
The situation is drastically different for single and double
spin asymmetries with a transversely polarized proton. In this case both
functions, $A$ and $B$, contribute.
Really, in some models the  amplitudes  $A$ and $B$
can have a phase shift. As a result, there appears the single spin asymmetry
determined by the pomeron exchange:
\be
A^h_{\perp} = \frac{2m \sqrt{|t|} \Im (AB^{*})}{|B|^2}. \lab{epol}
\ee
Here the $|A|^2$ term is omitted in the denominator.
So, the knowledge of all spin
structures in the pomeron-proton vertex is important here.
The model \ci{zpc} predicts that the asymmetry at $|t| > 1GeV^2$ can be about
$10 \div 15\%$.

The form of quark-pomeron coupling may be not so simple as in (\ref{pmu}).
 The perturbative calculations \ci{gol-pl}  for this vertex give:
\be
V_{qqI\hspace{-1.1mm}P}^{\mu}(k,r)=\gamma_{\mu} u_0+2 m k_{\mu} u_1 +
2 k_{\mu}
/ \hspace{-2.3mm} k u_2 + i u_3 \epsilon^{\mu\alpha\beta\rho}
k_\alpha r_\beta \gamma_\rho \gamma_5+im u_4
\sigma^{\mu\alpha} r_\alpha.    \lab{ver}
\ee
Here $k$ is a quark momentum, $r$ is a momentum transfer.
The spin structure of the quark-pomeron vertex (\ref{ver}) is
drastically different from the standard one (\ref{pmu}).
Really, the terms
$u_1(r)-u_4(r)$ lead to the spin-flip at the quark-pomeron vertex in contrast
to the term proportional to $u_0(r)$.  The functions
$u_1(r) \div u_4(r)$ at large $r^2$ are not very small \ci{gol4}. Note that
the phenomenological  coupling $V_{qqI\hspace{-1.1mm}P}^{\mu}$  with
$u_0$ and $u_1$ terms was proposed in \ci{klen}.

The nature of appearance of the new structures in the pomeron coupling
(\ref{ver}) is similar, e.g., to an rising of the anomalous magnetic momenta
of a particle. Really, if we calculate the single gluon loop correction
to the standard vertex (\ref{pmu}) for the massless quark, we obtain
the following structure (momentum definitions are as in (\ref{ver}))
\be
\gamma_{\alpha}(/ \hspace{-2.3mm} k+/ \hspace{-2.3mm} r) \gamma_{\mu}
/ \hspace{-2.3mm} k \gamma^{\alpha} \simeq -2[2(/ \hspace{-2.3mm} k+
\frac{/ \hspace{-2.3mm} r}{2}) k^\mu+i \epsilon^{\mu\alpha\beta\rho}
k_\alpha r_\beta \gamma_\rho \gamma_5].
\ee
So, in addition to the $\gamma_\mu$ term we obtain immediately from the loop
diagram the contributions equivalent to $u_2$ and $u_3$ in (\ref{ver}).
Expression
(\ref{ver}) can be regarded as a model for the effective quark-pomeron
coupling because only  planar graphs  have been considered. However, we hope
that the nonplanar diagrams that was excluded from the analysis \ci{gol-pl}
give a sufficiently small contribution at moderate momenta transfer.

This new form of the pomeron--quark coupling should modify various spin
asymmetries in high energy diffractive reactions \ci{klen,golasy}. It was
found from the analysis of longitudinal double spin asymmetries that the
main contribution is connected with  $u_0$ and $u_3$ in (\ref{ver}).
 The axial-like term
$V^{\mu}(k,r) \propto  u_3(r) \epsilon^{\mu\alpha\beta\rho}
k_\alpha q_\beta \gamma_\rho \gamma_5$ is  proportional to the momentum
transfer $r$  and thus contributes only in diffractive reactions where
the pomeron
has a non-zero momentum transfer ($r^2=|t|$). So, this new $u_3(r)$ term
does not change the standard pomeron contribution to the proton structure
functions because here the pomeron momentum transfer is equal to zero.

The aim of this report is to study  the pomeron-coupling effects in
single spin asymmetry in $pp$ diffractive high $p_t$-jet production. Future
HERA-N facilities \ci{now} is a good place to perform such investigations.

\section{Spin asymmetry in  diffractive reactions}

Let us investigate single transverse spin asymmetry in the
$p\uparrow p \to p+Q \bar Q+X$ reaction.
The standard kinematical variables look as follows
\be
s=(p_i+p)^2,\; t=r^2=(p-p')^2,\;x_p=\frac{p_i(p-p')}{p_i p}.
\ee
This process is determined
at small $x_p$ by the diagram in Fig.1.

   \begin{figure}[ht]
      \hspace*{1.9cm}
      \epsfxsize=11cm
      \centerline{\epsffile{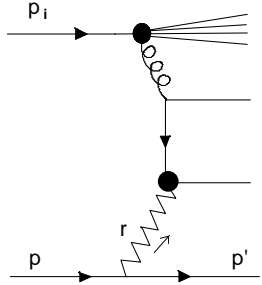}}
     \vspace*{-10.5cm}
   \caption{Diffractive $Q \bar Q$ production in $pp$ reaction. }
   \end{figure}
   \vspace*{.5cm}

The cross sections $\sigma$ and $\Delta \sigma$ can be written in the form
\be
\frac{d \sigma(\Delta \sigma)}{dx_p dt dp_{\perp}^2}=\{1,A^h_{\perp}\}
\frac{\beta^4 |F_p(t)|^2 \alpha_s}{128 \pi s x_p^2}
\int_{4p_{\perp}^2/sx_p}^{1} \frac{dy g(y)}{\sqrt{1-4p_{\perp}^2/syx_p}}
\frac{ N^{\sigma(\Delta \sigma)}
(x_p,p_{\perp}^2,u_i,|t|)}{(p_{\perp}^2+M_Q^2)^2}. \lab{si}
\ee
Here $g$ is the gluon structure function
of the proton, $p_{\perp}$ is a transverse momentum of jets, $M_Q$
is a quark mass, $N^{\sigma(\Delta \sigma)}$ is a trace over the quark loop,
$\beta$ is a pomeron coupling constant, $F_p$ is a pomeron-proton form factor.
In (\ref{si}) the coefficient equal to unity appears in $\sigma$ and
the transverse hadron asymmetry $A^h_{\perp}$ at the pomeron-proton vertex
determined in (\ref{epol}) appears in $\Delta \sigma$.

The main contributions to $N^\sigma(N^{\Delta \sigma})$
are determined by $u_0$ and $u_3$ structures in (\ref{ver}).
They can be written in the form for $x_p=0$:
\ba
N^{\Delta \sigma}=16(p_{\perp}^2+|t|) p_{\perp}^2 u_0^2+\Delta N^{\Delta
\sigma}; \nonumber\\
N^{\sigma}=32(p_{\perp}^2+|t|) p_{\perp}^2 u_0^2+\Delta N^{\sigma}.
\lab{nsy}
\ea
where
\ba
\Delta N^{\Delta\sigma}=8[(p_{\perp}^2+|t|)u_3-2u_0]
(p_{\perp}^2+|t|) p_{\perp}^2 |t| u_3; \nonumber \\
\Delta N^{\sigma}=16[(p_{\perp}^4+4p_{\perp}^2|t|+|t|^2)u_3-2
(2p_{\perp}^2+|t|) u_0] p_{\perp}^2 |t| u_3.\lab{a}
\ea
Note that $u_3<0$.

 Both $\sigma$ and $\Delta \sigma$ have a similar dependence at small $x_p$
$$ \sigma(\Delta \sigma) \propto \frac{1}{x_p^2} $$
This important property of (\ref{si}) allows one to study asymmetry at small
$x_p$ where the pomeron exchange predominates because a high energy
in the quark-pomeron system.

We shall calculate integrals (\ref{si}) using the simple form
for the gluon structure function

$$  g(y)=\frac{R}{y} (1-y)^5,\;\;\;\;R=3. $$
This form corresponds to the pomeron with $\alpha_{I\hspace{-1.1mm}P}(0)=1$.
Just the same approximation for the pomeron exchange has been used in
calculations. The analysis can be made for the  pomeron with
$\alpha_{I\hspace{-1.1mm}P}(0)=1+ \delta $ ($\delta>0 $) and more complicated
gluon structure functions but it does not change the results drastically.

In the diffractive jet production investigated here  the main contribution
is determined by the region where the quarks
in the loop are not far of the mass shell.
Then the interaction time should be long and the pomeron
rescatterings can be important. They change properties of single
pomeron exchange. This type of the pomeron is called usually the
"soft pomeron". It can have a spin-flip part with a phase different from
the spin-non-flip amplitude \ci{zpc}.
So, we can assume that the hadron asymmetry factor in (\ref{si}) can be
determined by the soft pomeron that coincides
with the elastic transverse hadron asymmetry
(\ref{epol}). In our further estimations we shall use the magnitude
$A^h_{\perp}=0.1$.

The calculations were performed for the magnitude $\beta=2GeV^{-1}$
\ci{don-la}
and the exponential form of the proton form factor
$$ |F_p(t)|^2=e^{bt} \;\;\;{\rm with}\;\; b=5GeV^2.$$

We used the
simple form of the $u_0(r)$  function:
$$   u_0(r)=\frac{\mu_0^2}{\mu_0^2+|t|} ,\;\;\;r^2=|t|,$$
with $\mu_0 \sim 1Gev$ introduced in  \ci{don-la}. The functions
$u_1(r) \div u_4(r)$ at $|t|>1 GeV^2$ were calculated in perturbative QCD
\ci{gol4}.

Our predictions for $\sigma$ and single spin asymmetry
 at the HERA-N energy $\sqrt{s}=40GeV$, $x_p=0.05$ and $|t|=1GeV^2$
for the standard quark-pomeron vertex (\ref{pmu}) and spin-dependent
vertex (\ref{ver}) are shown
in Figs.2 and 3 for light--quark jets.
It is easy to see that the shape of asymmetry is different for standard and
spin-dependent pomeron vertices. In the first case it
is approximately constant,
in the second it depends on $p^2_{\perp}$. This is caused by
the additional $p^2_{\perp}$ terms that appear in
$\Delta N^{\Delta \sigma}$ and $\Delta N^{\sigma}$ in (\ref{nsy}) for the
pomeron coupling (\ref{ver}).

   \begin{figure}[htb]
      \epsfxsize=13cm
      \centerline{\epsffile{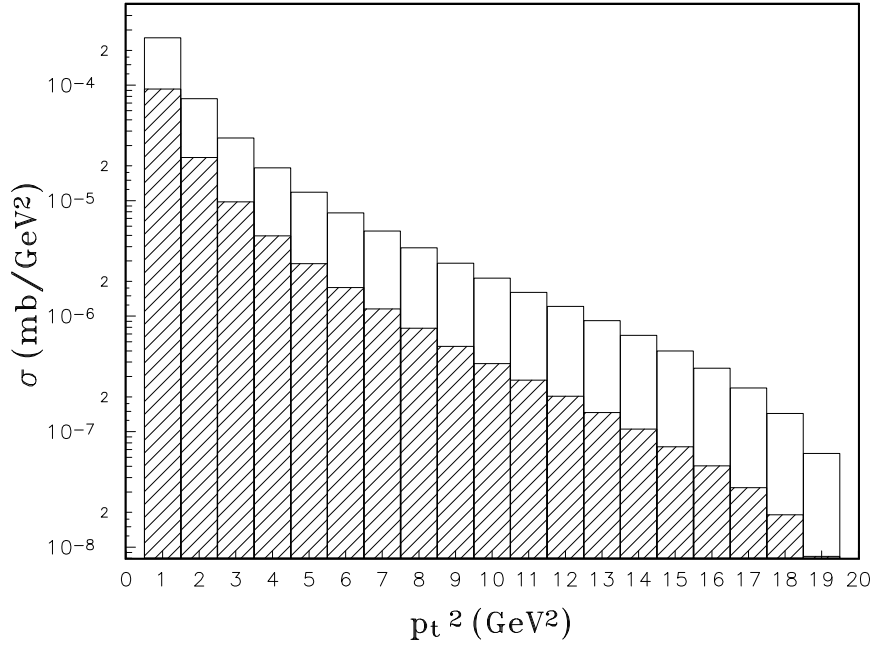}}
     \vspace*{-.5cm}
   \caption{$p^2_{\perp}$ dependence of $\sigma$. Fill
    boxes-for standard; open boxes -for spin-dependent quark-pomeron vertex.}
   \end{figure}
   \vspace*{.5cm}

To show the possibility to determine the form of the quark--pomeron coupling
>from experimental data, it is necessary to estimate possible errors.
For this purpose the experimental sensitivity in measured single spin
asymmetry obtained in \ci{now}
$$ \delta A \simeq \frac{0.1}{\sqrt{\sigma[pb]}} $$
was used. The estimated errors are shown in Fig.3. It easy to see that
the structure of the quark-pomeron vertex can be studied from the
$p^2_{\perp}$
distribution of single-spin asymmetry (the appropriate region is
$1Gev^2 <p^2_{\perp}< 10GeV^2$).

   \begin{figure}[ht]
      \epsfxsize=13cm
      \centerline{\epsffile{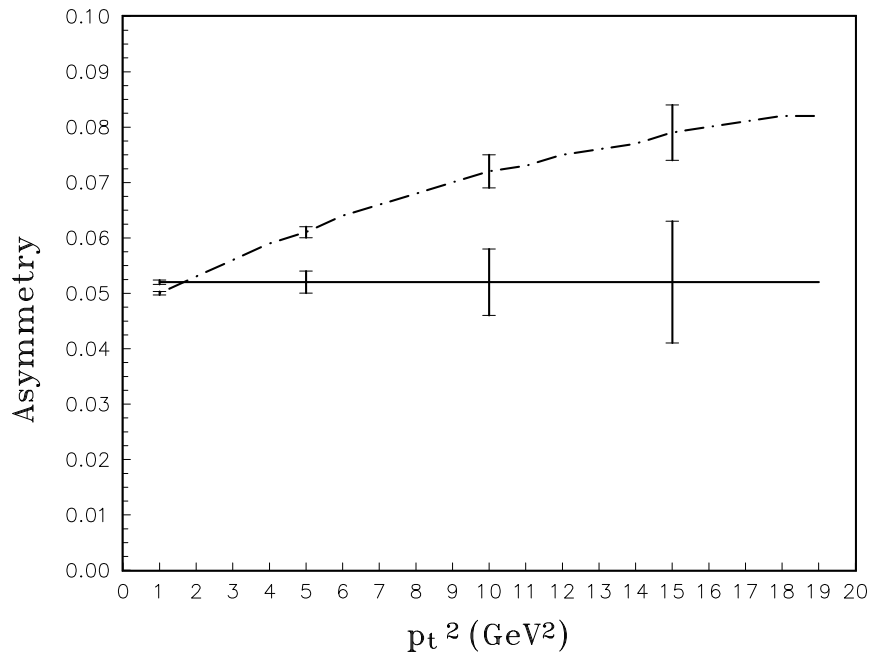}}
     \vspace*{-.5cm}
   \caption{$p^2_{\perp}$ dependence of asymmetry and the estimated
   errors. Solid line -for standard;
   dot-dashed line -for spin-dependent quark-pomeron vertex. }
   \end{figure}
  \vspace*{.5cm}

We calculate the cross sections $\sigma$ and $\Delta \sigma$ integrated
over $p^2_{\perp}$ of jets, too.
The asymmetry obtained from these integrated cross
sections does not depend practically on the quark-pomeron vertex structure.
It can be written in  both the cases in form
\be
A1=\frac{\int dp^2_{\perp} \Delta  \sigma}{\int dp^2_{ \perp}\sigma} \simeq
0.5 A^h_{\perp}  \lab{a1}
\ee
As a result, the integrated asymmetry (\ref{a1}) can be used for studying of
the hadron asymmetry $A^h_{\perp}$ caused by the pomeron.

The kinematics of the investigated reaction in the lab. system has been
studied. The recoil hadron should be emitted at an angle about
$40^o \div 60^o$. The typical jet angles should be $40 \div 100 mrad$. So,
they can be detected by HERA-N. To detect the recoil hadron it is
necessary to have the RECOIL detector for $\theta_{Lab} \simeq
40^o \div 60^o$.

Thus, in this report the perturbative QCD analysis of single spin asymmetry
in diffractive 2-jet production  in the $pp$ reaction is performed.
It is shown
that such experiments at HERA-N energies permit one to study spin
properties of quark-pomeron and proton-pomeron vertices determined
by QCD at large distances.

I am grateful to W.-D.Nowak for fruitful
discussion and hospitality at DESY-Zeuthen where this report has been
completed.

This work was supported in part by the Russian Foundation for
Fundamental Research,
Grant 94-02-04616, and Heisenberg-Landau Grant.


\normalsize
\end{document}